\documentclass[aps,10pt,prl,floatfix,footinbib,reprint]{revtex4-1}
\usepackage{graphicx,textcomp,dsfont,natbib,hyperref,color,amsmath}

\hypersetup{
  colorlinks   = true, 
  urlcolor     = blue, 
  linkcolor    = black, 
  citecolor   = blue 
}

\begin{document}
\title{Interferometric laser cooling of atomic rubidium}
\author{Alexander Dunning}\altaffiliation{Present address: UCLA Department of Physics \& Astronomy, 475 Portola Plaza, Los Angeles, CA 90095, USA}\email{alexander.dunning@gmail.com}
\author{Rachel Gregory}
\author{James Bateman}
\author{Matthew Himsworth}
\author{Tim Freegarde}
\affiliation{School of Physics \& Astronomy, University of Southampton, Highfield, Southampton SO17 1BJ, UK}
\date{\today}

\begin{abstract}
We report the 1-D cooling of $^{85}$Rb atoms using a velocity-dependent optical force based upon Ramsey matter-wave interferometry. Using stimulated Raman transitions between ground hyperfine states, 12 cycles of the interferometer sequence cool a freely-moving atom cloud from $21~\mu$K to $3~\mu$K. This pulsed analog of continuous-wave Doppler cooling is effective at temperatures down to the recoil limit; with augmentation pulses to increase the interferometer area, it should cool more quickly than conventional methods, and be more suitable for species that lack a closed radiative transition.
\end{abstract}
\maketitle

The laser cooling of atomic gases has revolutionized experimental atomic physics~\cite{Wieman1999Atom} and raised the prospect of a range of atomic quantum technologies~\cite{Dowling2003Quantum,Georgescu2012}. However, traditional Doppler  cooling~\cite{Hansch1975Cooling,Wineland1979Laser} relies upon the velocity-dependence of a single narrow radiative transition, and spontaneous emission to reset the atomic state. The cooling force is limited to a half photon-impulse per excited-state lifetime and, as many impulses are needed, requires a transition that can be closed by a few repump lasers. Doppler cooling has thus so far been limited to a handful of atomic elements and molecules~\cite{Shuman2010Laser,Hummon20132D,Lien2014Broadband,Zhelyazkova2014Laser}.

In \cite{Weitz2000Frequencyindependent}, Weitz and H\"{a}nsch proposed a mechanism that could extend laser cooling to a wider range of species, by replacing the continuous wave (CW) excitation of conventional Doppler cooling with the broadband laser pulses of Ramsey matter-wave interferometry, and interleaving inversion pulses to eliminate the dependence upon the internal state energies. The interference signal, and hence the impulse imparted, were thus determined only by the particle's kinetic energy; manifold transitions could be accessed and, while spontaneous emission remained the entropy-removing mechanism, various schemes~\cite{Bakos1996Transient,Freegarde2006Coherent,Freegarde2003Algorithmic} could increase the impulse per spontaneous event. With a drive towards efficient pulsed schemes for molecular cooling~\cite{Kielpinski2006Laser,Jayich2014Continuous,Romanenko2014Cooling} supported by improved mode-locked laser technologies, interferometric cooling appears a promising and flexible tool.

The idea of a pulsed Ramsey analog to CW Doppler cooling has until now remained untested. In this letter, we report the first experimental demonstration of 1-D interferometric cooling of a cloud of already ultracold Rb atoms. Our long-lived quasi-two-level system, comprising the two $5\mathrm{S}_{1/2}$ ground hyperfine states of $^{85}$Rb between which we drive stimulated Raman transitions, in principle allows cooling to the recoil limit, and we show that with just 12 cycles of the interferometric cooling sequence the atom cloud is cooled from $21\pm2~\mu$K to $3.2\pm0.4~\mu$K. Relaxation after each cycle is achieved by rapid pumping and decay of the single-photon $5\mathrm{S}_{1/2}$--$5\mathrm{P}_{3/2}$ transition, and the cooling rate is therefore limited mainly by the time needed for interferometric resolution of the different velocity classes.

\begin{figure}[t!]
  \centering
  \includegraphics[width=8.5cm]{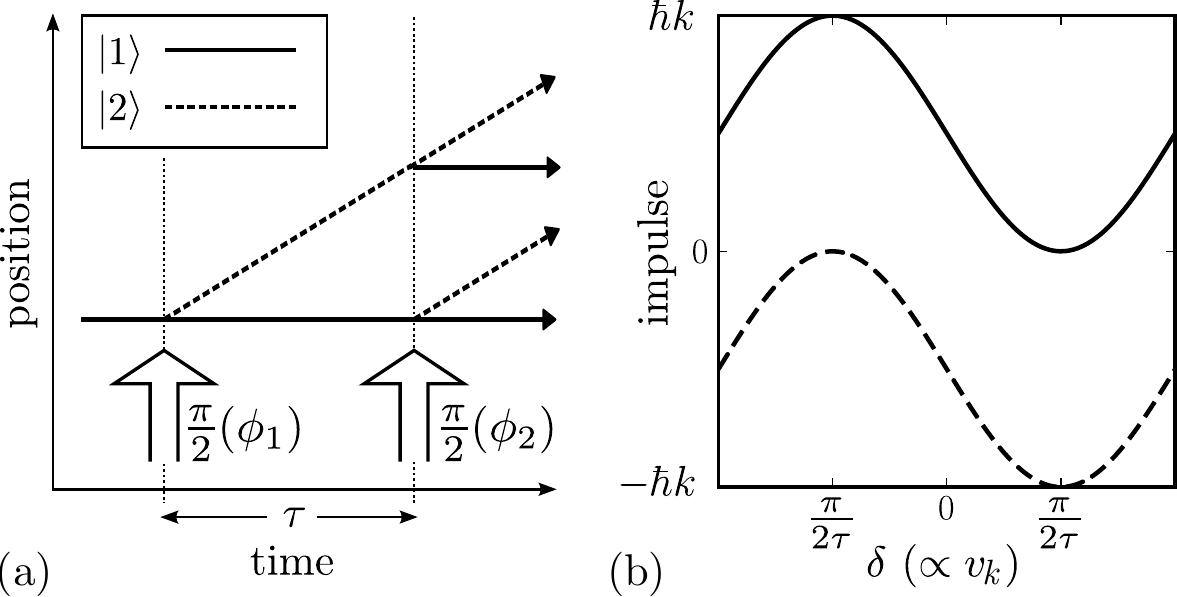}
  \caption{\label{Ramsey}(a) Position-time trajectory of a two-level atom in the velocity-selective Ramsey interferometer: two co-propagating $\pi/2$ pulses with phases $\phi_1$ and $\phi_2$, separated by a dwell time $\tau$, split and recombine the matter-wave. (b) Impulse imparted as a function of the detuning $\delta$ (proportional to the atom's velocity $v_k$) for positive (solid) and negative (dashed) wavevectors $\mathbf{k}$, with a relative phase $\phi_{\mathrm{rel}} = \phi_{1}-\phi_{2} = -\frac{\pi}{2}$.}
\end{figure}

\begin{figure*}[t!]
  \centering
  \includegraphics[width=17.8cm]{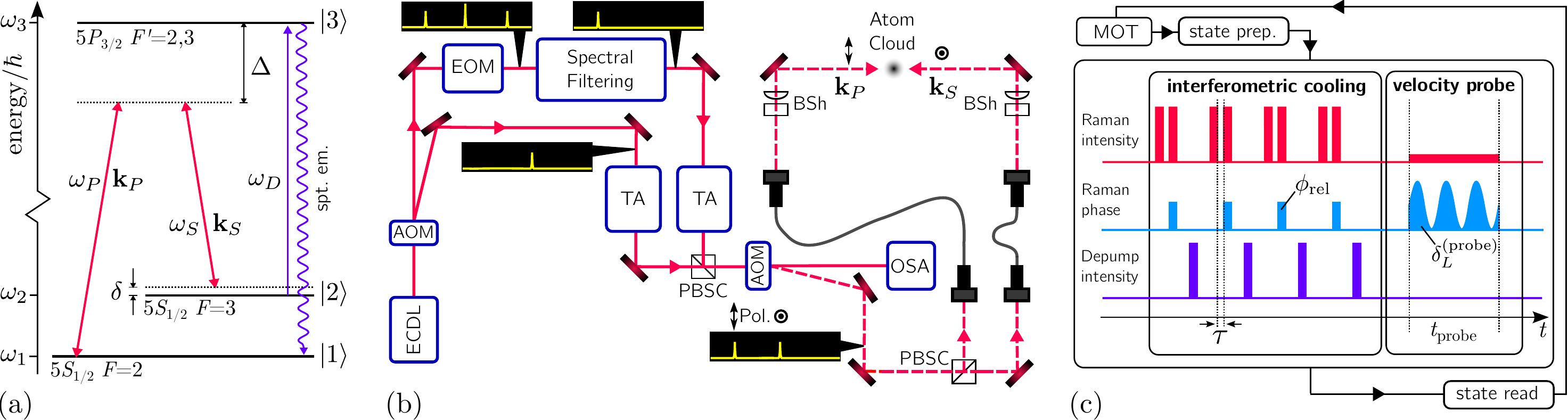}
  \caption{\label{Experiment}(a) Energy level diagram for the interferometric cooling experiment in $^{85}$Rb. (b) Schematic of the experimental setup of the Raman beams: ECDL -- external-cavity diode laser; PBSC -- polarizing beamsplitter cube; TA -- tapered amplifier; OSA -- optical spectrum analyzer; BSh -- beam shaper and focusing lens. The annotation bubbles show sketches of the beam spectrum at each preparation stage. (c) Experimental sequence diagram for $N=4$ applications of the cooling sequence.}
\end{figure*}

The Raman interferometric cooling mechanism is as follows. Two $\pi/2$ laser pulses, separated by a dwell time $\tau$, act upon a two-level atom $\vert\Psi \rangle = c_{1}\vert1\rangle + c_{2}\vert2\rangle$ as the beamsplitter and combiner of a Ramsey matter-wave interferometer, as illustrated in Fig.~\ref{Ramsey}(a). With the atom starting in $\vert1\rangle$, the final excitation probability
\begin{equation}\label{Eq1}
\vert c_{2} \vert^{2} \simeq \tfrac{1}{2} \big[1+\cos\left(\delta \, \tau - \phi_{\mathrm{rel}}\right)\big]
\end{equation}
depends upon the detuning $\delta$ of the pulses from the rest-frame resonance, the phase difference $\phi_{\mathrm{rel}}=\phi_{1}-\phi_{2}$ between them, and the dwell time $\tau$. The detuning
\begin{equation}\label{DeltaEq}
\delta = \delta_{L} + \mathbf{k}\cdot\left(\mathbf{v}+\mathbf{v}_{R}/2 \right)
\end{equation}
depends upon the component of the initial atom velocity $\mathbf{v}$ along the laser wavevector $\mathbf{k}$, the detuning $\delta_{L}$ at zero velocity, and the recoil velocity $\mathbf{v}_{R}$. Since excitation to  $\vert2\rangle$ changes the atom's momentum by $m \mathbf{v}_{R} = \hbar \mathbf{k}$, the mean impulse, shown in Fig.~\ref{Ramsey}(b), has a sinusoidal dependence on the velocity. The fringes can be shifted in velocity by changing $\phi_{\mathrm{rel}}$; their period can be increased by reducing $\tau$; and the impulse can be reversed by reflecting $\mathbf{k}$ or starting the sequence in state $\vert2\rangle$. Just as the Doppler effect renders photon absorption more probable for atoms moving towards a red-detuned CW laser beam, light-pulse Ramsey interferometry can thus impart a velocity-dependent impulse through the Doppler-shifted interference pattern.

When the two-level interferometer is implemented by counterpropagating Raman laser pulses with frequencies $\omega_{P,S}$ and wavevectors $\mathbf{k}_{P,S}$ along the $z$-axis, the impulse $\hbar\mathbf{k}_{\mathrm{eff}} = \hbar(\mathbf{k}_{P}-\mathbf{k}_{S})$ is of magnitude $\hbar k_{\mathrm{eff}}\approx 2\hbar k$, where $k \approx \vert \mathbf{k}_{P} \vert \approx \vert \mathbf{k}_{S} \vert$ is the single photon wavenumber; and the Raman detuning from the rest-frame atomic resonance is $\delta_{L} = (\omega_{P}-\omega_{S})-(\omega_{1}-\omega_{2}+\delta_\textnormal{AC})$, where $\hbar \omega_{1,2}$ are the atomic state energies and $\delta_{AC}$ is the combined AC Stark shift in the presence of the beams. If $\delta_{L}=0$, or the dependence upon it is eliminated as in \cite{Weitz2000Frequencyindependent} by including appropriately timed inversion pulses, the detuning depends only upon the velocity component $v_{k} \equiv \mathbf{v}\cdot\mathbf{k}_{\mathrm{eff}}/\vert\mathbf{k}_{\mathrm{eff}}\vert$ of the atom. We thus have the convenience of a large impulse and r.f. frequency difference, and can use the long-lived ground hyperfine states.

Although a single interferometer sequence alters the atomic velocity distribution, it does not itself increase the phase space density, because the decelerated atoms are excited to a different internal state. Dissipation is achieved through relaxation by spontaneous emission to the initial state. Whereas this in the initial conception occurs naturally~\cite{Weitz2000Frequencyindependent}, the spontaneous emission rate in our Raman scheme is negligible; we therefore end the interferometer sequence by pumping atoms from state $\vert2\rangle$ to the 5P$_{3/2}$ state, labelled $\vert3\rangle$ in Fig.~\ref{Experiment}(a), from where spontaneous emission returns them to the lower states with a lifetime of 26~ns~\cite{SteckRubidium}. This switchable relaxation allows the hyperfine distribution to be reset promptly without limiting the coherence time of the interferometer.

For cooling to occur, the interferometer phase $\phi_{\mathrm{rel}}$ is set to give either a negative slope $\mathrm{d}\vert c_{2}\vert^{2}/\mathrm{d}v_{k}$ across the velocity distribution for an initial state $\vert1\rangle$, or a positive slope for atoms starting in state $\vert2\rangle$. The velocity dependent impulse is, as noted by Weitz and H\"{a}nsch~\cite{Weitz2000Frequencyindependent}, in each case accompanied by a velocity-independent impulse, which may be cancelled by alternating between these two combinations, or by exchanging the directions of the counterpropagating Raman beams and hence $\mathbf{k}_{\mathrm{eff}}$.

The velocity capture range $\Delta v_{k} = \Delta \delta/k_{\mathrm{eff}} = \pi/k_\mathrm{eff}\tau$, within which the impulse increases monotonocally with velocity, is given by the width $\Delta\delta = \pi/\tau$ of the negative slope region of Fig.~\ref{Ramsey}(b). For a cloud of $^{85}$Rb atoms with temperature $T=100\,\mu$K  and hence r.m.s. velocity of $\sigma_{v} = \sqrt{k_{\mathrm{B}}T/m} = 0.1\,\mathrm{m\,s}^{-1}$, and 780~nm Raman beams, the dwell time $\tau$ required to capture the velocity distribution up to $\pm3\sigma_{v}$ is roughly 330~ns. Alternatively, for a sample at $T=1$~K, this falls to around $\tau=3$~ns.

Our experimental sequence, illustrated in Fig.~\ref{Experiment}(c), is as follows. $^{85}$Rb atoms from a background gas are initially trapped and cooled in a standard 3-D magneto-optical trap (MOT). The magnetic fields are then extinguished, the beam intensities reduced, and the cloud left to thermalize in the 3-D molasses for 5~ms. Because atoms at the centre of the molasses undergo sub-Doppler cooling more readily than those at the edges~\cite{Townsend1995Phasespace}, the velocity distribution at this point exhibits a two-component Gaussian shape, with half the population in a central peak at a temperature of around $20\,\mu$K and the rest in a broader background above the Doppler cooling limit ($146\,\mu$K) at around $250\,\mu$K.

The MOT repumping laser, which is resonant with the $5S_{1/2}\,F=2\rightarrow5P_{3/2}\,F=3$ transition, is then extinguished, and the atoms are optically pumped in $300\,\mu$s into the $5S_{1/2}\,F=2$ ground hyperfine state by the MOT cooling laser, which is red-detuned from the $5S_{1/2}\,F=3\rightarrow5P_{3/2}\,F=4$ transition. Three mutually orthogonal sets of shim coils cancel the residual magnetic field at the cloud position, such that the Zeeman sub-levels $m_{F}=-F\ldots F$ for each hyperfine state are degenerate to much less than the Rabi frequency $\Omega_{\mathrm{eff}}\approx 2\pi\times400$~kHz observed for the Raman transition.

\begin{figure*}[t!]
  \centering
  \includegraphics[width=17.8cm]{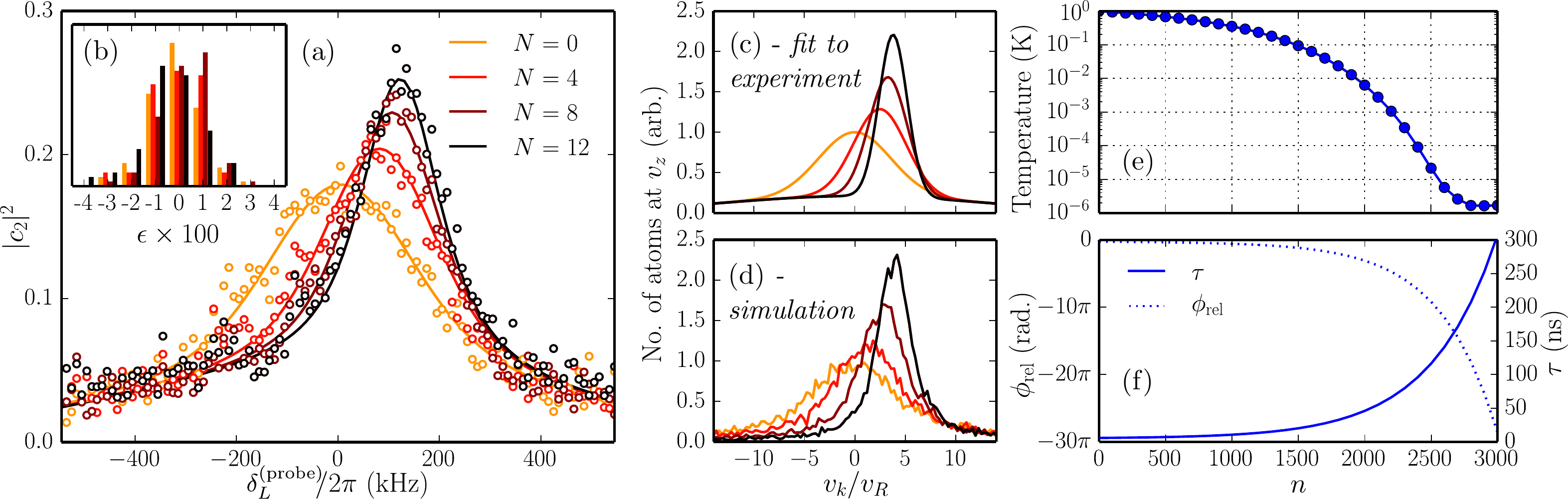}
  \caption{\label{Results}(a)~Raman velocimetry measurements after $N$ interferometric cooling cycles. Each point is an average of 16 shots at probe detuning $\delta_L^{(\mathrm{probe})}$, and the lines are from manually-fitted numerical simulations of the probe pulse assuming different velocity distributions. (b)~Histogram of the residuals $\epsilon = |c_2|^2_\textnormal{fit}-|c_2|^2_\textnormal{exp}$ corresponding to panel (a). (c) Velocity distributions corresponding to the solid curves in panel (a). (d) Velocity distributions from a numerical simulation of the cooling sequence with $\tau = 1.1~\mu$s and $\phi_\textnormal{rel}=-3\pi/8$. (e) Simulated 1--D temperature, starting at 1~K, vs. number of cooling cycle applications $n$ for an example scheme in which $\tau$ and $\phi_\textnormal{rel}$ vary with $n$ as illustrated in panel (f).}
\end{figure*}

The $\pi/2$ interferometer pulses are realized by driving stimulated Raman transitions between the $5S_{1/2}$ $F=2$ and $F=3$ ground hyperfine levels, using 780~nm beams, with $\delta_L = 0$ and detuned from the $5P_{3/2}$ states, as illustrated in Fig.~\ref{Experiment}(a). Each Raman transition lasts a quarter Rabi cycle, and is too rapid to resolve velocities within the ultracold sample. Spontaneous emission is induced by optical pumping from $5S_{1/2}\,F=3$ into $5P_{3/2}\,F'=3$, using a `depumping' laser, denoted $\omega_{D}$ in Fig.~\ref{Experiment}(a), aligned perpendicular to $\mathbf{k}_{\mathrm{eff}}$. The interferometer-depump sequence is applied $N$ times before the velocity distribution is measured using Raman velocimetry, whereby a long ($t_\textnormal{probe}=200~\mu$s), weak Raman pulse excites into $5S_{1/2}\,F=3$ a narrow velocity class defined by the probe pulse detuning~\cite{Reichel1994Subrecoil} according to Equation~\ref{DeltaEq}. The number of atoms in $5S_{1/2}\,F=3$ is then measured by light-induced fluorescence (LIF) excited by the MOT cooling laser, and the result is normalised to the total atom number by immediately repumping the whole distribution into $5S_{1/2}\,F=3$, and repeating the LIF measurement. The velocity distribution is determined by repeating the sequence at a range of probe detunings.

The source of our Raman pulses is shown schematically in Fig.~\ref{Experiment}(b). The continuous-wave beam from a 780~nm external cavity diode laser, red-detuned from single-photon resonance by $\Delta\approx2\pi\times11$\,GHz, is spatially divided by a 310~MHz acousto-optical modulator (AOM), and the rest of the microwave frequency shift is achieved by passing the undeflected beam through a 2.726~GHz electro-optical modulator (EOM). We control the EOM phase and frequency using an in-phase and quadrature-phase (IQ) modulator, fed from a pair of arbitrary waveform generators. The carrier wave is removed after the EOM using a polarizing beamsplitter cube~\cite{Cooper2012Actively}, and temperature-dependent birefringence within the EOM is countered by active feedback to a liquid crystal phase retarder~\cite{Bateman2010HanschCouillaud}. The remaining off-resonant sideband is removed using a stabilized fibre-optic Mach--Zehnder interferometer~\cite{Cooper2013Stabilized}.

After pre-amplifying the EOM sideband by injection-locking a CW diode laser, the two spectrally pure Raman beams are individually amplified by tapered laser diodes, recombined with orthogonal polarizations and passed through an AOM (rise time $\sim370~$ns), whose first-order output forms the Raman pulse beams. The beams are then separated by a polarizing beamsplitter, and passed via optical fibres to the MOT chamber.

After the fibres, each beam is passed through a Topag GTH-4-2.2 refractive beam shaper and 750~mm focal length lens to produce an approximately uniform $1.4$~mm square beam whose intensity varies by $\lesssim15$\% across the MOT cloud. Each beam has an optical power of 50~mW, and hence an intensity around $2.5\, \mathrm{W\,cm}^{-2}$ -- significantly higher than the large-waist Gaussian beams required for the same spatial homogeneity. To avoid broadening effects due to sublevel-dependent light shifts, the Raman beams have orthogonal linear polarizations. For the cooling sequence, the dwell time $\tau$ and relative phase $\phi_{\mathrm{rel}}$ were nominally set to $600$~ns (which, allowing for the AOM rise time, becomes $970$~ns) and $-\pi/2$, respectively.

Our experimental results for $N=0, 4, 8$ and 12 consecutive applications of the interferometric cooling cycle are shown in Fig.~\ref{Results}(a), where the circles show the measured excited state populations $|c_2|^2$ after the probe pulse at a range of probe detunings. These four data sets were acquired interleaved within the same experimental run to ensure comparable conditions, and the probe detunings were sampled pseudo-randomly to counter any effects of experimental drift. 
To estimate the temperatures, we fit numerical simulations (solid lines), for a common set of model parameters (beam intensities, detunings, timings, phases etc.), varying the two-component velocity distributions whose free parameters are the width and central $v_k$ of the colder Gaussian component: we assume that a fraction of 0.49 of the atom cloud remains at the background temperature of $250~\mu$K throughout. As a qualitative indicator of the fit quality, Fig.~\ref{Results}~(b) shows a histogram of the residuals $\epsilon = |c_2|^2_\textnormal{fit}-|c_2|^2_\textnormal{exp}$ for the range $-400$~kHz~$ < \delta_L/2\pi < 400$~kHz, revealing an approximately normal distribution. The corresponding velocity distributions are plotted in Fig.~\ref{Results}(c). The initial, uncooled distribution ($N=0$) exhibits a central Gaussian component at $21\pm2~\mu$K, which after $N=4$ cooling cycles this component has cooled to $10\pm1~\mu$K. For $N=8$ the temperature is $4.8\pm0.5~\mu$K, and after $N=12$, the temperature has reached $3.2\pm0.4~\mu$K.
The Raman recoil temperature of this system -- our theoretical lower temperature limit -- is $1.5\,\mu$K.

In our experiment, all interferometer pulses carry the same $\mathbf{k}_{\mathrm{eff}}$, hence as $N$ increases the atoms are accelerated towards positive velocity. Weitz and H\"ansch noted that population can be made to accumulate around $v_{k}=0$ by alternating between interferometer impulses $\pm\hbar\mathbf{k}_{\mathrm{eff}}$; similarly, the scheme could be extended to 3-D by introducing interferometer impulses along orthogonal axes~\cite{Weitz2000Frequencyindependent}.

To validate our experimental results, we have numerically simulated the cooling scheme using ProtoMol~\cite{Matthey2004Object}. An impulse $\Delta\mathbf{p} = \hbar\mathbf{k}_{\mathrm{eff}}\frac{1}{2}[1+A\cos(\delta\tau - \phi_\textnormal{rel})]$, followed by a randomly directed spontaneous emission recoil, is applied $N$ times to an ensemble of 10,000 atoms. The parameter $A=0.8$ describes the fringe contrast, and is experimentally measured using an adaptation of the four-pulse Ramsey-Bord\'{e} interferometer \cite{Borde1989Atomic}, used in previous photon recoil measurements~\cite{Weiss1993Precision}. The resultant velocity distributions for an initial central temperature of $20~\mu$K, and taking $\tau = 1.1~\mu$s and $\phi_\textnormal{rel}=-3\pi/8$, are plotted in Fig.~\ref{Results}(d), and clearly resemble our experimental results. The differences from the nominal values of $\tau$ and $\phi_\textnormal{rel}$ can be attributed to the finite AOM rise time, and its combination with uncompensated IQ delay lines, respectively.

We also use a numerical simulation to predict the performance at higher temperatures. For efficient cooling, the dwell time $\tau$ and phase $\phi_{\mathrm{rel}}$ must evolve as the velocity distribution narrows and shifts~\cite{Weitz2000Frequencyindependent}. For the $n$th cooling cycle of the example shown in Figs.~\ref{Results}(e,f), $\tau = \tau_0(1+ae^{bn})$ and $\phi_\textnormal{rel} = \phi_0 + n \tau v_R k_\mathrm{eff} / 2$, where $\tau_0 = 5~$ns, $a = 0.15$, $b = 2\times10^{-4}$ and $\phi_0 = -\pi/4$. As illustrated in Fig.~\ref{Results}(e) where we begin at 1~K, the temperature falls roughly exponentially with the number of interferometer sequences $n$, before arriving at the recoil limit at $n\sim 2800$. With this non-optimised dynamic scheme, about 65\% of the atoms remain within the Gaussian distribution as the recoil limit is reached.

For a given radiative transition, interferometric and CW Doppler cooling have similar cooling rates and limiting temperatures, determined by the radiative lifetime. However, reducing the interferometer period $\tau$ allows a larger capture range for higher temperatures without saturating the transition for lower velocities, while Raman transitions allow sub-Doppler temperatures to be reached without the usual Sisyphus mechanisms.

As our Rabi frequency is limited by the available intensities, we have not yet implemented the inversion pulses of the scheme's original proposal~\cite{Weitz2000Frequencyindependent}, which remove the dependence of cooling on the detuning of the laser from atomic resonance; our cooling scheme therefore relies upon a single atomic transition. However, improvements in beam intensity, spectral control and switching speed should allow interferometric cooling on multiple atomic transitions simultaneously. Considerably higher intensities have been demonstrated~\cite{Chiow2012Generation} which could raise the Rabi frequency $\Omega_{\mathrm{eff}}$ above $2\pi\times$100~MHz, allowing the full interferometer sequence and capture ranges up to 1~K. For the broad bandwidth originally envisaged for the cooling of molecules distributed across rotational manifolds, a rather different approach using mode-locked lasers and spectral beam shaping~\cite{Weiner2000Femtosecond} would be needed.

A variety of other enhancements could extend and improve this cooling scheme. Adiabatic chirps~\cite{Goldner1994Momentum,Bateman2007Fractional} and composite pulses~\cite{Levitt2007,Dunning2014Composite} allow interferometer fidelity to be extended over more complex pulse sequences and greater systematic inhomogeneities. These in turn permit amplified~\cite{Bakos1996Transient,Freegarde2006Coherent} and algorithmic~\cite{Freegarde2003Algorithmic} cooling techniques that offer faster cooling and, by reducing the number of spontaneous emission events accompanying a given cooling impulse, may be of particular value with more complex spectra within which population will inevitably decay into dark states not addressed even with the broadest of laser pulses.

This work was supported by the UK Engineering and Physical Sciences Research Council (Grant No. GR/S71132/01).

\providecommand{\noopsort}[1]{}\providecommand{\singleletter}[1]{#1}%

\end{document}